# CRYOMDOULE TEST STAND REDUCED-MAGNETIC SUPPORT DESIGN AT FERMILAB*

M.W. McGee[†], S.K. Chandrasekaran, A.C. Crawford, E. Harms, J. Leibfritz, G. Wu
Fermi National Accelerator Laboratory, Batavia, IL 60510, USA

*Abstract*

In a partnership with SLAC National Accelerator Laboratory (SLAC) and Jefferson Lab, Fermilab will assemble and test 17 of the 35 total 1.3 GHz cryomodules for the Linac Coherent Light Source II (LCLS-II) Project. These devices will be tested at Fermilab's Cryomodule Test Facility (CMTF) within the Cryomodule Test Stand (CMTS-1) cave. The problem of magnetic pollution became one of major issues during design stage of the LCLS-II cryomodule as the average quality factor of the accelerating cavities is specified to be 2.7 x $10^{10}$. One of the possible ways to mitigate the effect of stray magnetic fields and to keep it below the goal of 5 mGauss involves the application of low permeable materials. Initial permeability and magnetic measurement studies regarding the use of 316L stainless steel material indicated that cold work (machining) and heat affected zones from welding would be acceptable.

## INTRODUCTION

The Cryomodule Test Facility (CMTF) at Fermilab is a research and development facility for accelerator science and technology, in particular, the testing and validating of Superconducting Radio Frequency (SRF) components. CMTF provides the necessary test bed to measure and characterize the performance of SRF cavities in a cryomodule. CMTF was designed to be a flexible test facility, configurable in different ways to meet the needs of current as well as future projects at Fermilab and abroad [1].

The purpose of CMTS-1, is to test cryomodules of various frequencies in pulsed or continuous wave mode. It is currently being prepared to support the testing of cryomodules for the LCLS-II project being built at Stanford Linear Accelerator (SLAC). It will test both 1.3 and 3.9 GHz cryomodules in Continuous Wave (CW) mode for LCLS-II [1].

Fermilab is responsible for testing seventeen 1.3 GHz and two 3.9 GHz cryomodules which are of a modified TESLA-style design to accommodate the higher heat load inherent of CW operation and LCLS-II beam parameters. Each cryomodule consists of eight superconducting RF cavities, a magnet package, and instrumentation including a Beam Position Monitor. A total of thirty-five of these cryomodules, approximately half built at Fermilab and half at Jefferson Lab, will become the main accelerating elements of the 4 GeV Linac. The modifications and special features of the cryomodule include: thermal and cryogenic design to handle high heat loads in CW operation, magnetic shielding and cool-down configurations to enable high quality factor (Q0) performance of the cavities and liquid helium management to address the different liquid levels in the 2-phase pipe with 0.5% SLAC tunnel longitudinal slope [2].

## CRYOMODULE DESIGN

The LCLS-II 1.3 GHz cryomodule shown in Figure 1 consists of eight dressed 9-cell niobium superconducting radio frequency (RF) cavities with a quadrupole and dipole corrector package found at the downstream (DS) end. This coldmass hangs from three column support posts constructed from G-10 fiberglass composite, which are attached to the top of the vacuum vessel. The helium gas return pipe (HeGRP), supported by the three columns act as the coldmass spine, supporting the cavity string and ancillaries. Support brackets with adjusting blocks using needle bearings on each side provide a connection between each cavity and the HeGRP. An aluminum 45 K heat shields hang from the HeGRP column supports with a 5 K thermal intercept [2].

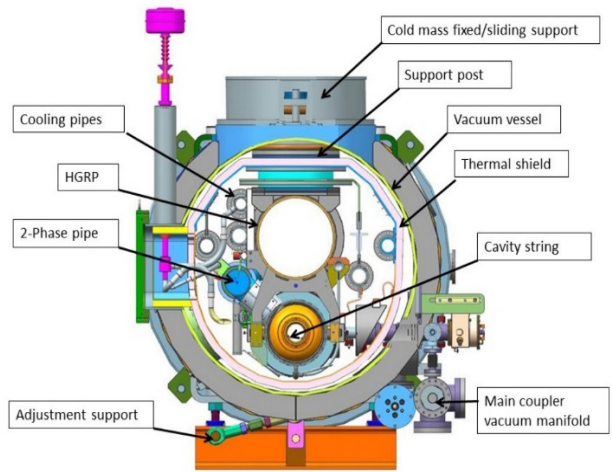

Figure 1: Cross-section of the cryomodule showing its major sub-assemblies [2].

## CMTS-1 CAVE

The CMTS-1 test cave is a shielded enclosure sized to house cryomodules as large as TESLA-style 8 cavity 1.3 GHz ones. Inner dimensions are 19.74 m long by 4.57 m wide with a height of 3.2 m. The walls are composed of shielding blocks and are 0.91 m thick with integrated penetrations for RF waveguide, cabling, etc. The roof is removable in order to move cryomodules in and out of the cave and is similarly composed of blocks with a total thickness of 0.91 m [3]. Figure 2 shows a cryomodule assembly attached to the CMTS-1 girder within the cave at CMTF.

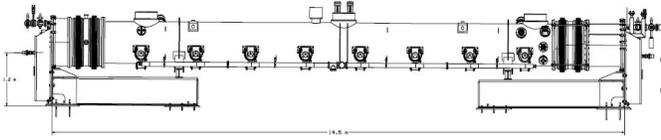

Figure 2: CMTS-1 girder assembly within cave.

## *Cryomodule Support Girders*

The original cryomodule girder design was conceived by INFN, Milan and used at DESY for XFEL. This design was later modified slightly to accommodate the use of available English structure steel shapes and standard AISC sections [4] while maintaining important metric standard features such as feed/end cap interface hole placement. The modified version of the DESY girder design was applied at Fermilab's New Muon Lab (NML) to support both cryomodules, CM1 and CM2, using feed/end caps manufactured by Cryotherm [5].

Recent findings show that the cavity quality factor, $Q_0$ is enhanced by expelling the magnetic flux during rapid cooldown. Meeting the LCLS-II $Q_0$ specification of 2.7 x $10^{10}$ requires an ambient magnetic field of 5 mGauss or less, averaged over the RF surface of the cavities [6,7]. Stray magnetic fields external to the cryomodule may be more than the assumed average and must be minimized.

There are two ways to approach the stray field issue (or minimize the effect); provide magnetic shielding or eliminate the source(s). Each girder consists of common structural steel found either in Europe (used by DESY) or within the USA (applied here at Fermilab). The chosen approach involved changing the original girder material specification from A36 structural steel to 316L stainless steel. This change also applies to any hardware used within the design, especially near the cryomodule. Structural steel hardware (such as critical girder and Hilti ™ bolt connections) used with the original design have been retained in the CMTS-1 girder design. These components are far enough from the cavity string to prevent stray field exposure.

The basic structure consists of two pseudo w-beams (0.61 m in height) welded together. Roughly, 3.1 m from the feed cap end there is a vacuum vessel support with a hole-pattern to receive the cryomodule adjustable stand and bears one half of the cryomodule weight; 3,688 kg. At the other end, the two top and bottom W-beam flanges are coped to accept two top and bottom plates which are welded to the webs of the two horizontal beams. Between these two top and bottom plates, there are two vertical gusset plates used to compensate for the moment produced by the vacuum load, an 111,600 N reaction. Then, at the same end, on top, the top plate and triangle gussets are welded directly above the two plates. Two pseudo W-beams (406 mm height x 38.1 mm thick web and flanges) support the cryomodule end plate and is attached to the end assembly described above and shown in Figure 3. The Feedcap and Endcap girders are essentially identical in terms of the structural integrity, therefore only one analysis is considered.

The CMTS-1 test stand simulates the 0.5% slope (based on the Earth's curvature) found within the Linac portion of the SLAC tunnel as shown in Figure 3.

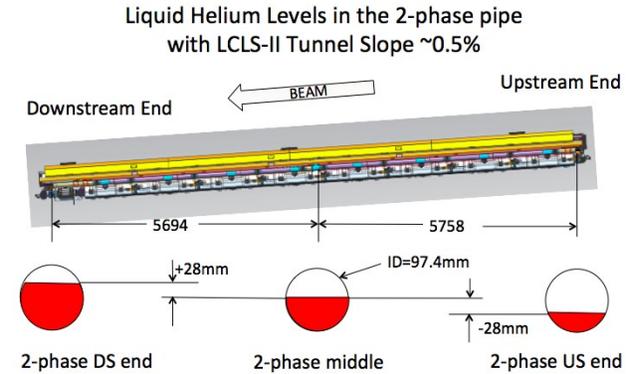

Figure 3: Liquid levels in the 2-phase pipe with 0.5% SLAC tunnel slope, with pipe sized to accommodate liquid levels both downstream and upstream (US) [2].

## MATERIAL AND CONSTRUCTION

Pre-design testing of material and the fabrication processes provided insight regarding the possible outcome of the stand fabrication. First, the cold work (machining) effect was evaluated by considering the machined feature with the most material removed and area affected. Initial cold work permeability was < 1.25, following drilling a 25.4 mm diameter hole; the highest stray remnant field measured was < 350 mGauss. Initial fillet and butt weld samples had permeability of 1.10, following the welding. The permeability increased to 2.0 in both cases. Post welding remnant fields reached ~350 mGauss.

Both a set of cryomodule girders and stands for CMTS-1 were fabricated based on our initial understanding of possible issues regarding remnant magnetic fields. The design material choice was essentially 316L stainless steel with a few exceptions. Grade 8 bolts were required given the high vacuum load reaction of 111,600 N.

The CMTS-1 adjustable stands were intended as a prototype for the LCLS-II Linac cryomodule stand design. Fermilab is responsible for the stand design. Since the LCLS-II cryomodules attach from the floor and seismic activity must be factored, there are differences. The adjustability of the stand design is identical in both cases. Changing the materials to primarily stainless steel had implications to the stand functionally. Initial testing of the CMTS-1 stand revealed difficulties making adjustments due to increased friction of the stainless steel components involved. The supporting rod and bearing disc pairs required the use of (carbon) steel and subsequent hardening as was previously implemented on CM-1 and CM2.

Applying material choices such as 316L stainless steel throughout the Fermilab LCLS-II stand design will become cost prohibitive at a point given a total of 70 required stands and two spares. Therefore, low carbon steel may be used conditionally for LCLS-II installation. In either case, stainless or low carbon steel, post fabrication (machining and welding) magnetic measurement is required. Drawings and other fabrication documents must specify any remnant field to be less than or equal to 3 Gauss at contact as an acceptance QC step. The basis of this specification considers that the H field at a 152 mm distance will decay to near earth magnetic field, if the measured contact magnetic field is < 3 Gauss. At each support location, a cavity is ~ 300 mm away from the cryomodule support stands. Finally, low carbon steel can be demagnetized if needed.

## MAGNETIC MEASURMENT

The magnetic field along the beamline and of components in CMTS-1 were measured to determine if the girder design and fabrication went as planned. These shall be used as 'baseline' magnetic field distribution of the CMTS-1 area.

A three-axis magnetic field sensor was used to measure the fields. The field on-contact with the components around the cryogenic feed and end cap and the support girder were measured to be less than 100 mGauss for most components, with a cryogenic line measuring in excess of 500 mGauss above background. Steel clamps used on the outer body connect points of the end caps also displayed magnetic fields greater than 500 mGauss. Although high on-contact, the field from these decayed rapidly with distance and did not seem to have a significant impact on the field distribution at the cavity's surface. Figure 4 provides the magnetic field measurements at CMTS-1, where the measurement was split into two sections, one starting from the US end (black and green) and one from the DS end (red and blue).

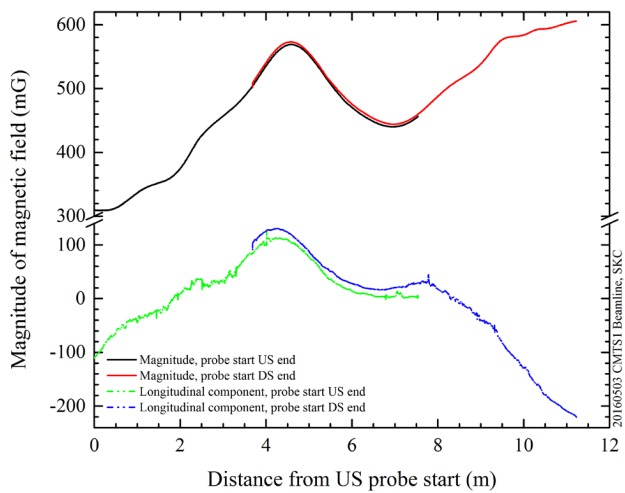

Figure 4: The measured magnetic field, magnitude (solid lines) and longitudinal component (dash-dot lines), along the beamline in CMTS-1.

The longitudinal component, for the most part, is below 150 mGauss. It is therefore, roughly the same as the fields in the SLAC tunnel.

## FUTURE WORK

The LCLS-II 1.3 GHz and 3.9 GHz cryomodule supporting stand design was been completed by Fermilab with the CMTS-1 experience in mind.

## ACKNOWLEDGEMENTS

We wish to thank LCLS-II Cryomodule Fermilab Project Managers; Camille Ginsberg, Rich Stanek and Jay Theilacker. Thanks to the SRF Technical Group members; Craig Rogers and Wayne Johnson for their necessary assistance. Also, thanks to the Fermilab Alignment and Metrology Group (AMG) staff: Virgil Bocean, Gary Crutcher, Mike O'Boyle, Gary Teafoe and Chuck Wilson.